\documentclass[11pt]{article}
\usepackage{amsmath,amssymb,amsthm}
\usepackage{url}
\usepackage{pgf}
\usepackage{epic}
\usepackage{eepic}
\usepackage{latexsym}
\usepackage{mathrsfs}
\usepackage{graphics}

\newtheorem{theorem}{Theorem}[section]
\newtheorem{lemma}[theorem]{Lemma}

\newtheorem{proposition}[theorem]{Proposition}

\newtheorem{definition}[theorem]{Definition}

\newcommand{\pc}{{\rm pc}}

\usepackage{xspace}

\usepackage{vmargin}

\setmarginsrb{3.8cm}{3.9cm}{3.9cm}{3.9cm}{0.5cm}{0.5cm}{0.5cm}{0.5cm}

\newcommand{\2}{\vspace{0.2 cm}}
\newcommand{\tw}{\mbox{\normalfont tw}}

\title{Minimum Leaf Out-branching and Related
Problems\thanks{Preliminary extended abstract of this paper appears
in the proceedings of AAIM'08 \cite{GuRaKi}.}}

\author{
 Gregory Gutin\thanks{Department of Computer Science,
 Royal Holloway, University of London,
Egham, Surrey TW20 0EX, UK, {\tt gutin@cs.rhul.ac.uk}} \and Igor
Razgon\thanks{Department of Computer Science, University College
Cork, Ireland, {\tt i.razgon@cs.ucc.ie}} \and Eun Jung
Kim\thanks{Department of Computer Science,
 Royal Holloway, University of London,
Egham, Surrey TW20 0EX, UK, {\tt eunjung@cs.rhul.ac.uk}} }

\begin{document}
\date{}
\maketitle

\begin{abstract}
\noindent Given a digraph $D$, the Minimum Leaf Out-Branching
problem (MinLOB) is the problem of finding in $D$ an out-branching
with the minimum possible number of leaves, i.e., vertices of
out-degree 0. We prove that MinLOB is polynomial-time solvable for
acyclic digraphs. In general, MinLOB is NP-hard and we consider
three parameterizations of MinLOB. We prove that two of them are
NP-complete for every value of the parameter, but the third one is
fixed-parameter tractable (FPT). The FPT parametrization is as
follows: given a digraph $D$ of order $n$ and a positive integral
parameter $k$, check whether $D$ contains an out-branching with at
most $n-k$ leaves (and find such an out-branching if it exists). We
find a problem kernel of order $O(k^2)$ and construct an
algorithm of running time $O(2^{O(k\log k)}+n^6),$ which is an
`additive' FPT algorithm. We also consider transformations from two
related problems, the minimum path covering and the maximum internal
out-tree problems into MinLOB, which imply that some
parameterizations of the two problems are FPT as well.
\end{abstract}

\section{Introduction}\label{introsec}

We say that a subgraph $T$ of a digraph $D$ is an {\em out-tree} if
$T$ is an oriented tree with only one vertex $s$ of in-degree zero
(called {\em the root}). The vertices of $T$ of out-degree zero are
called {\em leaves} and all other vertices {\em internal vertices}. If $T$
is a spanning out-tree, i.e. $V(T)=V(D)$, then $T$ is called an {\em
out-branching} of $D$. Given a digraph $D$, the {\em Minimum Leaf
Out-Branching} problem ({\em MinLOB}) is the problem of finding an
{out-branching} with the minimum possible number of leaves in $D$. We denote this minimum by $\ell_{\min}(D)$ and if $D$ has no
out-branching, we write $\ell_{\min}(D)=0$.  Notice that not every
digraph $D$ has an out-branching. It is not difficult to see that
$D$ has an out-branching (i.e., $\ell_{\min}(D)>0$) if and only if
$D$ has just one strongly connected component without incoming arcs
\cite{bang2000}. Since the last condition can be checked in linear
time \cite{bang2000}, we may often assume that $\ell_{\min}(D)>0$.

The {\em underlying graph} $UG(D)$ of a digraph $D$ is obtained from
$D$ by omitting all orientation of arcs and by deleting one edge
from each resulting pair of parallel edges. For a digraph $D$, an independent set (vertex cover respectively) of $D$ is an independent set (vertex cover respectively) of $UG(D)$. We denote the union of (in-) out-neighbors of vertices in $X$ by ($N^-(X)$) $N^+(X)$ and $N(X)=N^+(X)\cup N^-(X)$.

We first study MinLOB restricted to acyclic digraphs (abbreviated
{\em MinLOB-DAG}). MinLOB-DAG was considered in US patent
\cite{demers2000}, where its application to the area of database
systems was described. Demers and Downing \cite{demers2000} also
suggested a heuristic approach to MinLOB-DAG. However no argument or
assertion has been made to provide the validity of their approach
and to investigate its computational complexity. Using another
approach, we give a simple proof in Section \ref{dagsec} that
MinLOB-DAG can be solved in polynomial time.

Since MinLOB generalizes the hamiltonian directed path problem,
MinLOB is NP-hard. In this paper, we introduce three
parameterizations of MinLOB: (a) $\ell_{\min}(D)\le k$ ($k\ge 1$),
(b) $\ell_{\min}(D)\le n/k$ ($k\ge 2$), (c) $\ell_{\min}(D)\le n-k$
($k\ge 1$), where $n$ is the number of vertices in $D$ and $k$ is
the parameter. We show that (a) and (b) are NP-complete for every
value of the parameter, but (c) is fixed-parameter tractable and has an algorithm of complexity $O(2^{O(k\log k)}+n^6)$. We also show the existence of a quadratic kernel for the parameterized problem (c). These results are considered in section (\ref{parsec})-(\ref{solsec}). The problem (c) was studied by Prieto and Sloper \cite{prieto2003, prieto2005} for undirected graphs (i.e. symmetric digraphs), where the authors obtained an FPT algorithm of complexity $O(2^{2.5k\log k}n^O(1))$ and a quadratic kernel.

In the {\em minimum path covering} problem ({\em MinPC}), given a
digraph $D$, our aim is to find the minimum number of
vertex-disjoint directed paths, $\pc(D)$, covering all vertices of
$D$. It is well-known that MinPC is polynomial time solvable for
acyclic digraphs, $\pc(D)\le \alpha(D)$ for every digraph $D$ (the
Gallai-Milgram theorem), and  $\pc(D)= \alpha(D)$ for every
transitive acyclic digraph $D$ (Dilworth's theorem) \cite{bang2000}.
In first part of Section \ref{outsec}, we describe a simple
transformation from MinPC into MinLOB which implies that the
parameterized problem $\pc(D)\le n-k$ is fixed-parameter tractable,
where $n$ is the number of vertices in $D$ and $k$ is the parameter.

Observe that MinLOB can be reformulated as a problem of finding an
out-branching with maximum number of internal vertices. However, while the
problem of finding an out-tree with minimum number of leaves in a
digraph $D$ is trivial (a vertex is an out-tree), the problem of
{\em finding an out-tree with maximum number of internal vertices}
(abbreviated {\em MaxIOT}) is not trivial; in fact, it is NP-hard
(as it also generalizes the hamiltonian directed path problem). The
second part of Section \ref{outsec} is devoted to the latter
problem. Further research is discussed in Section \ref{ressec}.

We recall some basic notions of parameterized complexity here, for a
more in-depth treatment of the topic we refer the reader to
\cite{downey1999,flum2006,niedermeier2006}.

A parameterized problem $\Pi$ can be considered as a set of pairs
$(I,k)$ where $I$ is the \emph{problem instance} and $k$ (usually an
integer) is the \emph{parameter}.  $\Pi$ is called
\emph{fixed-parameter tractable (FPT)} if membership of $(I,k)$ in
$\Pi$ can be decided in time $O(f(k)|I|^c)$, where $|I|$ is the size
of $I$, $f(k)$ is a computable function, and $c$ is a constant
independent from $k$ and $I$. Let $\Pi$ be a parameterized problem. A \emph{reduction $R$ to a problem kernel} (or \emph{kernelization}) is a many-to-one transformation from $(I,k)\in\Pi$ to $(I',k')\in \Pi'$, such that (i) $(I,k)\in \Pi$ if and only if $(I',k')\in \Pi$, (ii) $k'\le k$ and $|I'|\le g(k)$ for some function $g$ and (iii) $R$ is computable in time polynomial in $|I|$ and $k$. In kernelization, an instance $(I,k)$ is
reduced to another instance $(I',k')$, which is called the
\emph{problem kernel}; $|I'|$ is the {\em size} of the kernel.

It is easy to see that a decidable parameterized problem is FPT if
and only if it admits a kernelization (cf.
\cite{flum2006,niedermeier2006}); however, the problem kernels
obtained by this general result have impractically large size.
Therefore, one tries to develop kernelizations that yield problem
kernels of smaller size. The survey of Guo and Niedermeier
\cite{guoACMSN38} on kernelization lists some problem for which
polynomial size kernels and exponential size kernels were obtained.
Notice that due to kernelization we can obtain so-called an {\em additive
FPT} algorithm, i.e., an algorithm of running time
$O(n^{O(1)}+g(k)),$ where $g(k)$ is independent of $n$,
which is often significantly faster than its `multiplicative'
counterpart.

All digraphs in this paper are finite with no loops or parallel
arcs. We use terminology and notation of \cite{bang2000}; in
particular, for a digraph $D$, $V(D)$ and $A(D)$ denote its vertex
and arc sets. The symbols $n$ and $m$ will denote the number of vertices and arcs in the digraph
under consideration.

\section{MinLOB-DAG}\label{dagsec}

Let $D$ be an acyclic digraph. We may assume that $D$ has a unique vertex $r$ of in-degree 0
as otherwise $D$ has no out-branchings. Let $V=V(D)$ and $V'=\{v':\ v\in V\}.$
Let us define a bipartite graph $B$ of $D$ with partite sets $X$ and $X'$ as
follows: $X=V$, $X'=V'\setminus \{r'\}$ and $E(B) = \{xy':x\in X, y'\in X', xy \in
A(D)\}$.

Consider the following algorithm for finding a minimum leaf
out-branching $T$ in an input acyclic digraph $D$. The algorithm
outputs $T$ if it exists and `NO', otherwise.

\begin{quote}
  \noindent{\bfseries  MINLEAF}\\
 \begin{enumerate}
 \item[1.] {\bf if} the number of vertices with in-degree 0 equals 1 {\bf then}
\newline $r \leftarrow$ the vertex of in-degree 0 {\bf else} return `NO'
 \item[2.] construct the bipartite graph $B$ of $D$
 \item[3.] find a maximum matching $M$ in $B$
 \item[4.] $M^*\leftarrow M$
 \item[5.] {\bf for} all $y'\in X'$ not covered by $M$ {\bf do}
\newline $M^* \leftarrow M^*\cup \{$an arbitrary edge incident with $ y'\}$
 \item[6.] $A(T)\leftarrow \emptyset$
 \item[7.] {\bf for} all $xy' \in M^*$ {\bf do} $A(T)\leftarrow A(T)\cup\{xy\}$
 \item[8.] return $T$
\end{enumerate}
\end{quote}

\begin{theorem}\label{ACYminleaf}
Let $D$ be an acyclic digraph. Then MINLEAF returns a minimum leaf
out-branching if one exists, or returns `NO' otherwise in time
$O(m+n^{1.5}\sqrt{m/\log n})$.
\end{theorem}
\begin{proof} We start with proving the validity of the algorithm. Observe that an
acyclic digraph has an out-branching if and only if there exists
only one vertex of in-degree zero. Hence Step 1 returns `NO'
precisely when $\ell_{\min}(D)=0.$

Let $M$ be the maximum matching obtained in Step 2, let $V(M)$ be the set of vertices
of $B$ covered by $M$, and let $Z=X\setminus V(M)$ and $Z'=X'\setminus V(M).$

First we claim that $Z$ is the set of the leaves of $T$, the
out-branching of $D$ obtained in the end of Step 7. Consider the
edge set $M^*$ obtained at the end of Step 5. First observe that for
each vertex $y' \in Z'$, there exists an edge of $E(B)$ which is
incident with $y'$ since $r$ is the only vertex of in-degree zero
and thus no vertex of $Z'$ is isolated. Moreover, all neighbors of
$y'$ are covered by $M$ due to the maximality of $M$. It follows
that $M^*\supseteq M$ covers all vertices of $X'$ and leaves $Z$
uncovered. Notice that $r$ is covered by $M$. Indeed there exists a
vertex $u$ such that $r$ is the only in-neighbor of $u$ in $D$.
Hence if $r$ was not covered by $M$ then $u'$ would  not be covered
by $M$ either, which means we could extend $M$ by $ru'$, a
contradiction.

Consider $T$ which has been obtained in the end of Step 7. Clearly
$d^-_{T}(v)=1$ for all $v\in V(D)\setminus \{r\}$ due to the
construction of $M^*$. Moreover $D$ does not have a cycle, which
means that $T$ is connected and thus is an out-branching. Finally no
vertex of $Z$ has an out-neighbor in $T$ while all the other
vertices have an out-neighbor. Now the claim holds.

Conversely, whenever there exists a minimum leaf out-branching $T$
of $D$ with the leaf set $Z$, we can build a matching in $B$ which
covers exactly $X\setminus Z$ among the vertices of $X$. Indeed,
simply reverse the process of building an out-branching $T$ from
$M^*$ described at Step 7. If some vertex $x\in X$ has more than one
neighbor in $X'$, eliminate all but one edge incident with $x$.

Secondly we claim that $T$ obtained in MINLEAF($D$) is of minimum number of leaves.
Suppose to the contrary that the the attained out-branching $T$ is not a minimum
leaf out-branching of $D$. Then a minimum leaf out-branching can be used to produce a
matching of $B$ that covers more vertices of $X$ than $M$ does using the argument in
the preceding paragraph, a contradiction. Hence MINLEAF(D) returns a min leaf
out-branching $T$ at Step 8.

Finally we analyze the computational complexity of MINLEAF($D$).
Each step of MINLEAF($D$) takes at most O($m$) time except for Step
3. The computation time required to perform Step 3 is the same as
that of solving the maximum cardinality matching problem on a
bipartite graph. The last problem can be solved in time
$O(|V(B)|^{1.5}\sqrt{|E(B)|/\log |V(B)|})$ \cite{A}.  Hence, the
algorithm requires at most $O(m+n^{1.5}\sqrt{m/\log n})$ time.
\end{proof}

\section{Parameterizations of MinLOB}\label{parsec}

The following is a natural way to parameterize MinLOB.

\begin{quote}
  \noindent{\bfseries  MinLOB Parameterized Naturally (MinLOB-PN)}\\
  \emph{Instance:} A digraph $D$.\\
  \emph{Parameter:} A positive integer $k$.\\
  \emph{Question:} Is $\ell_{\min}(D)\le k$ ?
\end{quote}

Clearly, this problem is NP-complete already for $k=1$ as for $k=1$
MinLOB-PN is equivalent to the hamiltonian directed path problem.
Let $v$ be an arbitrary vertex of $D$. Transform $D$ into a new
digraph $D_k$ by adding $k$ vertices $v_1,v_2,\ldots ,v_k$ together
with the arcs $vv_1,vv_2,\ldots ,vv_k$. Observe that $D$ has a
hamiltonian directed path terminating at $v$ if and only if
$\ell_{\min}(D_k)\le k$. Since the problem is NP-complete of
checking whether a digraph has a hamiltonian directed path
terminating at a prescribed vertex, we conclude that MinLOB-PN is
NP-complete for every fixed $k$.

Clearly, $\ell_{\min}(D)\le n-1$ for every digraph $D$ of order
$n>1$. Consider a different parameterizations of MinLOB.

\begin{quote}
  \noindent{\bfseries MinLOB Parameterized Below Guaranteed Value (MinLOB-PBGV)}\\
  \emph{Instance:} A digraph $D$ of order $n$ with $\ell_{\min}(D)>0.$\\
  \emph{Parameter:} A positive integer $k$.\\
  \emph{Question:} Is $\ell_{\min}(D)\le n-k$ ?\\
  \emph{Solution:} An out-branching $B$ of $D$ with at
most $n-k$ leaves or the answer `NO' to the above question.
\end{quote}

Note that we consider MinLOB-PBGV as a search problem, not just as a
decision problem. In the next section we will prove that MinLOB-PBGV
is fixed-parameter tractable. We will find a problem kernel of order
$O(k\cdot 2^k)$ and construct an additive FPT algorithm of running
time $O(2^{O(k\log k)}+n^3).$ To obtain our results we use notions
and properties of vertex cover and tree decomposition of underlying
graphs and Las Vergnas' theorem on digraphs.

The parametrization MinLOB-PBGV is of the type {\em below a
guaranteed value}. Parameterizations above/below a guaranteed value
were first considered by Mahajan and Raman \cite{mahajanJA31} for
the problems Max-SAT and Max-Cut; such parameterizations have lately
gained much attention, cf.
\cite{fernau2005,gutinTCS41,gutinA,heggSTOC2007,niedermeier2006} (it
worth noting that Heggernes, Paul, Telle, and Villanger
\cite{heggSTOC2007} recently solved the longstanding minimum
interval completion problem, which is a parametrization above
guaranteed value). For directed graphs there have been only a couple
of results on problems parameterized above/below a guaranteed value,
see \cite{bang,fernauM2005}.

Let us denote by $\vec{K}_{1,p-1}$ the {\em star digraph} of order
$p$, i.e., the digraph with vertices $1,2,\ldots, p$ and arcs
$12,13,\ldots ,1p$. Our success with MinLOB-PBGV may lead us to
considering the following stronger (than MinLOB-PBGV)
parameterizations of MinLOB.

\begin{quote}
  \noindent{\bfseries MinLOB Parameterized Strongly Below Guaranteed
  Value \newline
(MinLOB-PSBGV)}\\
  \emph{Instance:} A digraph $D$ of order $n$ with $\ell_{\min}(D)>0.$\\
  \emph{Parameter:} An integer $k\ge 2$.\\
  \emph{Question:} Is $\ell_{\min}(D)\le n/k$ ?
\end{quote}

Unfortunately, MinLOB-PSBGV is NP-complete for every fixed $k\ge 2.$
To prove this consider a digraph $D$ of order $n$ and a digraph $H$
obtained from $D$ by adding to it the star digraph $\vec{K}_{1,p-1}$
on $p=\lfloor n/(k-1)\rfloor$ vertices ($V(D)\cap
V(\vec{K}_{1,p-1})=\emptyset$) and appending an arc from  vertex $1$
of $\vec{K}_{1,p-1}$ to an arbitrary vertex $y$ of $D$. Observe that
$\ell_{\min}(H)=p-1 + \ell_{min}(D,y)$, where $\ell_{min}(D,y)$ is
the minimum possible number of leaves in an out-branching rooted at
$y$, and that $\frac{1}{k}|V(H)|=p+\epsilon,$ where $0\le \epsilon
<1$. Thus, $\ell_{\min}(H)\le \frac{1}{k}|V(H)|$ if and only if
$\ell_{min}(D,y)=1.$ Hence, the hamiltonian directed path problem
with fixed initial vertex (vertex $y$ in $D$) can be reduced to
MinLOB-PSBGV for every fixed $k\ge 2$ and, therefore, MinLOB-PSBGV
is NP-complete for every $k\ge 2.$

\section{Quadratic Kernel for MinLOB-PBGV}\label{kernel}

In this section we introduce a reduction rule for the MinLOB-PBGV problem. Using the reduction rule we present a polynomial time algorithm that either yields an out-branching with at most $n-k$ leaves or produces a kernel whose size is bounded by a quadratic function of $k$.

Let $T$ be an out-branching of a given digraph $D$ and let $(u,v)\in A(D)\setminus A(T)$. We define the {\em 1-change} for $(u,v)$ as the operation to add the arc $(u,v)$ to $T$ and remove the existing arc $(p(v),v)$ from $T$, where $p(v)$ is the {\em parent} (i.e. in-neighbor) of $v$ in $T$. We say an out-branching is {\em minimal} if no 1-change for an arc of $A(D)\setminus A(T)$ leads to an out-branching with more internal vertices, or equivalently, less leaves. For distinct vertices $x,y$, we write $x<_T y$ if there is a path from $x$ to $y$ in $T$. An arc $(y,x)\in A(D)\setminus A(T)$ is $T$-backward if $x<_T y$. The following is a simple observation on a minimal out-branching.

\begin{lemma}\label{minimal}
Let $T$ be an out-branching of $D$. Then $T$ is minimal if and only if for every arc $(u,v)\in A(D)\setminus A(T)$ which is not $T$-backward arc, the vertex $u$ is internal or $d^+(p(v))=1$.
\end{lemma}

\begin{proof}
Suppose the 1-change for $(u,v)\in A(D)\setminus A(T)$ yields an out-branching with less leaves. It is easy to see that $(u,v)$ is not $T$-backward, $u$ is a leaf and $d^+(p(v))\geq 2$. Conversely if there is an arc $(u,v)\in A(D)\setminus A(T)$ which is not $T$-backward, $u$ is a leaf and $d^+(p(v))\geq 2$ then 1-change for $(u,v)$ produces an out-branching in which the number number of leaves is strictly decreased.
\end{proof}

\begin{lemma}\label{boundedvc}
Given a digraph $D$, we can either build a minimal out-branching $T$ with at most $n-k$ leaves or obtain a vertex cover of size at most $2k-2$ in $O(n^2m)$ time.
\end{lemma}
\begin{proof}
Let $T$ be a minimal out-branching. If $T$ has at most $n-k$ leaves, we are done. Suppose it is not. We claim that the set $U=\{u\in V(D):u$ is internal in $T\}\cup \{u\in V(D): u$ is a leaf in $T$ and $d^+(p(u))=1\}$ is a vertex cover of $D$. Since the set of internal vertices cover all arcs which are not between the leaves, it suffices to show that every arc $(u,v)$ between two leaves $u$ and $v$ is covered by $U$. The last statement follows from the fact that $T$ is minimal and Lemma \ref{minimal}. What remains is to observe that the number of internal vertices is at most $k-1$ and the number of leaves which is the only child of its parent is at most $k-1$ as well.

Now we consider the time complexity of the algorithm. The construction of an out-branching $T$ of $D$ takes $O(n+m)$ time. Whether $T$ is minimal can be checked in $O(nm)$ time since for every arc $(u,v)\in A(D)\setminus A(T)$ we test the conditions of Lemma \ref{minimal}. Let $L$ be the list of arcs $(u,v)\in A(D)\setminus A(T)$ which violates the minimality of $T$, i.e. such that $u$ is a leaf and $d^+(p(v))\geq 2$. Whenever $L\neq \emptyset$, choose $(u,v)\in L$ and transform $T$ by replacing the arc $(p(v),v)$ by $(u,v)$. Accordingly we update the list $L$ as follows: (1) erase all arcs whose tail is $u$, which takes $O(m)$ time (2) erase all arcs whose head is $v$, which takes $O(m)$ time (3) add to $L$ arcs of the form $(x,y)$ where $x$ is a leaf of the subtree rooted at $v$ and $y$ is a vertex with $d^+(p(y))\geq 2$ on the unique path from the root of $T$ to $p(v)$. This takes $O(nm)$ time. The validation of the update with (1)-(3) can be easily verified. Since any out-branching has at least one leaf and we decrease the number of leaves of $T$ by 1 at each transformation, after at most $n$ such transformations we obtain an out-branching where no further transformation can be done. This will be our minimal out-branching. When the minimal out-branching has more than $n-k$ leaves, we can construct the vertex cover $U$ as above in $O(n)$ time.
\end{proof}

It follows from Lemma \ref{boundedvc} that we can find either an out-branching which certifies a positive answer for the MinLOB-PBGV problem or a vertex cover of $D$ of size at most $2k-2$. In the second case, we can remove some redundant vertices from the large independent set of size at least $n-(2k-2)$ and obtain an instance of smaller size. The {\em crown structure} plays the fundamental role in this reduction.

\begin{definition}
A crown in a graph $G$ is a pair $(H,C)$, where $H\subseteq V(G)$ and $C\subseteq V(G)$ with $H\cap C=\emptyset$ such that the following conditions hold:

(a) The set of neighbors of vertices in $C$ is precisely $H$, i.e. $H=N(C)$,

(b) $C=C_m\cup C_u$ is an independent set, and

(c) There is a perfect matching between $C_m$ and $H$.
\end{definition}

A crown structure is a relatively new idea that allows us to have powerful reduction rules. Its applications have been wide and successful, which includes a linear-size kernel for the vertex cover problem \cite{chor2004, fellows2004}.\2

Given a digraph $D$, let $U$ be a vertex cover of $D$. Modify $U$ by including in it the vertex of in-degree 0 if one exists. Let $W=V(D)\setminus U$ and observe that $W$ is an independent set. Finding an out-branching with at most $n-k$ leaves can be reformulated as the problem of finding an out-branching with at least $k$ internal vertices. Herein we define the {\em internal number} of $D$ as the largest possible number of internal vertices of an out-branching of $D$.

In order to accommodate a crown structure to MinLOB-PBGV problem we create an auxiliary model which is similar to those considered in \cite{fellows2004, prieto2005}. Note that our model is more refined as we deal with directed graphs unlike \cite{fellows2004, prieto2005} which consider only undirected graphs. Given a directed graph $D$ with $U$ and $W$ as above, we build the (undirected) bipartite graph $B$ as follows.

\begin{itemize}
\item $V(B)=U'\cup W$, where $U'=N^{-}(W)\cup (U\times U)$
\item $E(B)=\{\{xy,w\}:xy\in U\times U, w\in W, (x,w)\in A(D), (w,y)\in A(D)\} \cup \{\{x,w\}:x\in U,w\in W, (x,w)\in A(D)\}$
\end{itemize}

Observe that no vertex of $W$ in $B$ is isolated since every vertex of $W$ is of in-degree at least one in $D$.

\begin{lemma}\label{reduction}
If $B$ contains a crown $(H,C=C_m\cup C_u)$ with $C\subseteq W$ and $C_u\neq \emptyset$, then the internal number of $D$ equals the internal number of $D- C_u$.
\end{lemma}
\begin{proof}
We can extend an out-branching $T$ of $D- C_u$ by appending an arc $(x,w)\in A(D)$, where $w\in C_u$ and $x$ is any in-neighbor of $w$. The attachment of such an arc does not decrease the number of internal vertices of $T$. This shows that the internal number of $D$ is not smaller than that of $D- C_u$.

Let a crown $(H,C=C_m\cup C_u)$ with $C\subseteq W$ and a perfect matching $M$ between $H$ and $C_m$ are given. We start with the following claim.\2

\noindent{\bf Claim 1.} \label{claim1} Let $c_{root}$ be the root of $T$. If $c_{root}\in C$, we can modify the perfect matching $M$ into $M'$ between $H$ and $C'_m\subseteq C$ so that $c_{root}\in C'_m$ and $\{ux,c_{root}\}\in M$ for some pair vertex $ux\in U\times U$. \2

\noindent{\em Proof of Claim 1.} Suppose this is not the case. Recall that $c_{root}$ is of in-degree at least 1 since we excluded any vertex of in-degree 0 from $W$. Let $u$ be an in-neighbor of $c_{root}$ in D and $x$ be a child of $c_{root}$ in $T$. Note that $\{u,c_{root}\},\{ux,c_{root}\}\in E(B)$ and thus $u,ux\in H$.


There are two cases and for each case we can obtain a new perfect matching as follows. Firstly if $c_{root}\in C_u$, simply exchange it with a vertex $c\in C_m$ which is matched to the pair vertex $ux$ by $M$. This exchange is justified since $\{ux,c_{root}\}\in E(B)$. Secondly suppose $c_{root} \in C_m$ but it is matched to a vertex $u\in N^-(W)$. Since $(u,c_{root}),(c_{root},x)\in A(D)$, we have the pair vertex $ux$ in $U'$ and moreover it is in $H$. Hence we can find $c\in C_m$ which is matched to the pair vertex $ux$ and by exchanging it with $c_{root}$ we have a new perfect matching. This is possible as we have $\{ux,c_{root}\} \in E(B)$ and $(u,c)\in A(D)$, thus $\{u,c\}\in E(B)$.\qed \2

Due to Claim 1, when $c_{root}\in C$ we may always assume that $c_{root}\in C_m$ and furthermore that $\{ux,c_{root}\}\in M$ for some pair vertex $ux \in (U \times U)$. Notice that $x$ is not necessarily a child of $c_{root}$ in $T$.

We shall show that the internal number of $D- C_u$ is not smaller than the internal number of $D$. To see this suppose $T$ is an out-branching of $D$ and consider the subgraph $F=T -C$ obtained from $T$ by deleting the vertices of $C$. Obviously $F$ is a union of out-trees, say $F_1,\ldots ,F_l$. We will add the vertices of $C_m$ and a set of arcs so that we obtain an out-branching of $D-C_u$ with as many internal vertices as in $T$ at the end of this process.

Recalling that $C\subseteq W$ is an independent set, it is straightforward to see any vertex $c\in C$ falls into one of the three types: (a) $c$ is a leaf in $T$ hanging to some vertex of $F$ (b) $c$ is an internal vertex in $T$ which has both a parent and children in $F$ (c) $c$ is the root $c_{root}$ of $T$ and it has at least one in-neighbor in $V(D)$.

Let $c_1,\ldots ,c_t \in C$ be the vertices that are of type (b) in $T$. Consider $c_i$, $1\leq i \leq t$. If $c_i$ comes under type (b), let $H_i=\{f_pf_{q}\in U\times U: (f_p,c_i)\in A(T), (c_i,f_q)\in A(T)\}$. We denote $\bigcup_{1\leq i\leq t}H_i$ by $H_{int}$. For the vertex $c_{root}\in C$, let $H_{c_{root}}=\{f_{p}x\in U\times U: (f_p,c_{root})\in A(D)\setminus A(T), (c_{root},x)\in A(T)\}$. We set $H_{root}=\emptyset$ if $c_{root} \notin C$. Note that both $H_{int}$ and $H_{root}$ belong to $H$.



The following procedure defines how to construct an out-tree $T''$ from $F$. We initialize $T'\leftarrow F$ and $C_{int}\leftarrow \emptyset$.

\begin{enumerate}
\item For every $f_pf_q \in H_{int}$
\newline 1.1 let $H_i$ be the unique set containing $f_pf_q$.
\newline 1.2 let $c_{pq} \in C_m$ be the vertex with $\{f_pf_q,c_{pq}\}\in M$
\newline 1.3 $T'\leftarrow T'+c_{pq}+(f_p,c_{pq})+(c_{pq},f_q)$.
\newline 1.4 $C_{int}\leftarrow C_{int} \cup c_{pq}$.

\item $T''\leftarrow T'$.

\item If $c_{root} \notin C$, return $T''$.

\item If $c_{root} \notin C_{int}$
\newline 4.1 $T''\leftarrow T'' + c_{root}$.
\newline 4.2  for each child $x$ of $c_{root}$ in $T$, $T''\leftarrow T''+(c_{root},x)$.
\newline 4.3 return $T''$.

\item Otherwise
\newline 5.1 let $f_pf_q\in H_{int}$ be the vertex with $\{f_pf_q,c_{root}\}\in M$.
\newline 5.2 let $x$ be the child of $c_{root}$ in $T$ with $x\leq_{T''} f_p$.
\newline 5.3 let $c_x\in C_m$ be the vertex with $\{f_px,c_x\} \in M$.
\newline 5.4 $T'' \leftarrow T'' + c_x + (c_x,x)$.
\newline 5.5 for each child $y\neq x$ of $c_{root}$ in $T$ (if any)
\newline 5.5.1 let $c_y\in C_m$ be the vertex with $\{f_py,c_y\}\in M$
\newline 5.5.2 $T''\leftarrow T''+ c_y +(f_p,c_y) + (c_y,y)$.
\newline 5.6 return $T''$
\end{enumerate}

\2
\noindent{\bf Claim 2.} \label{claim2} Step 1 is valid and $T'$ at step 2 is a union of out-trees.\2

\noindent{\em Proof of Claim 2.} For each $f_pf_q \in H_{int}$, the vertex $f_q\in V(F)$ appears as the second element of the pair vertex in $H_{int}$ at most once. The uniqueness of $H_i \ni f_pf_q$ then follows (step 1.1). Moreover by the construction of $H_i$, $\{f_pf_q, c_i\}\in E(B)$ and thus $f_pf_q \in N(C)=H$, where the last equality follows by the definition of crown. Hence $f_pf_q$ is uniquely matched to a vertex $c_{pq}\in C_m$ by $M$ (step 1.2). Also $\{f_pf_q, c_{pq}\}\in E(B)$ implies $(f_p,c_{pq}),(c_{pq},f_q) \in A(D)$, which implies that $T'$ can be properly constructed (step 1.3).

Now observe that any second element $f_q$ of a pair vertex $f_pf_q \in H_{int}$ is a root of an out-tree in $F$. Thus for each component $F_q$ of $F$, $T'$ contains at most one arc entering into its root. Moreover, $f_p<_{T'} f_q$ if and only if $f_p<_{T} f_q$, which means there is no directed cycle in $T'$. Witnessing that all the other vertices have at most one arc entering into it, we conclude $T'$ at step 2 is a union of out-trees. \qed \2

We claim that the above procedure returns an out-tree $T''$\2

\noindent{\bf Claim 3.} \label{claim3} Step 3-5 are valid and $T''$ is an out-tree.\2

\noindent{\em Proof of Claim 3.} First consider the case when $T''$ is returned at step 3. With Claim 2, it is enough to show that $T'$ is connected. Let two components $F_p$ and $F_q$ in $F$ be connected by $c_i$ in $T$. Since $c_{root}\notin C$, the vertex $c_i$ is of type (b) and thus there exist $f_p\in F_p$ and the root $f_q$ of $F_q$ such that $(f_p,c_i)\in A(T)$, $(c_i,f_q)\in A(T)$. By the construction of $H_{int}$, we have $f_pf_q\in H_i\subseteq H_{int}$ and the vertex $c_{pq}\in C_m$ with $\{f_pf_q,c_{pq}\}\in M$ connects $F_p$ and $F_q$ in $T'$ during the performance of step 1. Hence $T'$ is connected.

If $T''$ is {\em not} returned at step 3, we have $c_{root}\in C$. It is important to observe that in this case, the roots of the out-trees in $T'$ at step 2 are exactly the children of $c_{root}$ in $T$. This is because the root of an out-tree in $F$ has an incoming arc in $T'$ if and only if its parent in $T$ is of type (b).

Secondly suppose that $T''$ is returned at step 4. Then $c_{root}$ does not participate in $T'$ and $c_{root}$ in $T''$ is of in-degree 0. By the observation in the second paragraph, $T''$ is an out-tree.

Thirdly suppose that $T''$ is returned at step 5. In this case $c_{root}$ has been included as an internal vertex to connect two out-trees in step 1, and the arcs $(f_p,c_{root})$ and $(c_{root},f_q)$ have been included in $T'$, where $f_pf_q$ is the pair vertex found in step 5.1. We want to check that $c_x$ and the arc $(c_x,x)$ in line 5.3 can be properly picked up. Indeed, the pair vertex $f_px$ belongs to $H_{root}\subseteq H$ and there exists a vertex $c_x$ which is matched to the pair $f_px$. By the construction of $B$, the arc $(c_x,x)$ exists as well. Hence at the end of step 5.4, $T''$ is a union of out-trees whose roots are $c_x$ and the children of $c_{root}$ in $T$ other than $x$.

If $d^+_T(c_{root})=1$, $T''$ consists of a single out-tree whose root is $c_x$. Else if $d^+_T(c_{root})\geq 2$, let $y$ be a child of $c_{root}$ in $T$ and $y\neq x$. Since $(f_p,c_{root}), (c_{root},y) \in A(D)$, we have the pair vertex $f_py$ in $H_{root}\subseteq H$ and $f_py$ is uniquely matched to a vertex $c_y$. The edge $\{f_py,c_y\}$ implies the existence of the two arcs $(f_p,c_y)$, $(c_y,y)$, hence we can perform step 5.5 properly. Since the vertex $f_p$ is contained in the out-tree rooted at $c_x\in C_m$, the addition of these arcs does not create a cycle. As a result we start at the step 5.5 with $|d^+_T(c_{root})|$ out-trees in the beginning and each time we carry out step 5.5.2, the number of out-trees in $T''$ decreases by 1. Therefore at the end of step 5.5, we end up with a single out-tree $T''$ rooted at $c_x$.
\qed \2

During the construction of $T''$, we added at least one vertex $c_{pq}$ for each internal vertex $c_i$ of type (b) as an internal vertex of $T''$. Also we added at least one vertex as the root or an internal vertex of $T''$ if $c_{root}\in C$. Hence the number of internal vertices in $C$ for $T''$ is at least as large as the number of internal vertices in $C$ for $T$. Therefore what remains is to see that every vertex $f$ of $F$ which is internal in $T$ can be made to remain internal. The only case we need to consider is a vertex $f\in V(F)$ whose children in $T$ are leaves and all belong to $C$. Suppose $f$ is a leaf in $T'$. Since $f\in N(C)=H$, we can uniquely determine a vertex $c_f\in C_m$ such that $\{f,c_f\}$ belongs to the perfect matching $M$. By the construction of $T''$ in the above argument, the vertex $c_f$ is not contained in $T''$ for each such vertex $f\in V(F)$ and thus, we may add $c_f$ and an arc $(f,c_f)$ to $T''$ while keeping $T''$ as an out-tree. After this procedure each such vertex $f$ is an internal vertex in $T''$, and thus $T''$ has as many internal vertices as $T$.

For any vertex $c$ of $C_m$ which does not participate in $T'$ constructed so far, we simply add it to $T''$ with the arc $(f,c)\in A(D)$. Therefore $T''$ is an out-branching of $D- C_u$ with as many internal vertices as $T$. This completes the proof.
\end{proof}

In light of Lemma \ref{reduction}, we have a reduction rule below.

\2

\noindent{\bf Reduction rule 1.} Given a digraph $D$ with a vertex cover $U$ of $D$ and $W=V(D)\setminus U$, construct the associated bipartite graph $B$. If $B$ has a crown $(H,C=C_m\cup C_u)$ with $C_u\neq \emptyset$, remove the vertices of $C_u$ from $D$.\2

We need the following theorem to prove our kernelization lemma.

\begin{theorem}\cite{fellows2004}\label{crownexists}
Any graph $G$ with an independent set $I$, where $|I|\geq \frac{2n}{3}$, has a crown $(H,C)$, where $H\subseteq N(I)$, $C\subseteq I$ and $C_u\neq \emptyset$, that can be found in time $O(nm)$ given $I$.
\end{theorem}

\begin{lemma}[Kernelization Lemma] \label{kernellemma}
Let $D$ be irreducible. If $|V(D)|> 8k^2+6k$ then $D$ has an out-branching with at least $k$ internal vertices.
\end{lemma}
\begin{proof}
Suppose that $D$ is reduced with $|V(D)|> 8k^2+6k$, and that $D$ does not have an out-branching with at least $k$ internal vertices. Since the internal number of $D$ is the same as the internal number of the original digraph, we may assume that $D$ has an out-branching $T$.

For $|U|<2k$, we have $|W|=|V(D)\setminus U|> 8k^2+4k$ and $|U'|< 2k+4k^2$. Then $|W|\geq \frac{2|V(B)|}{3}$ which means we have a crown $(H,C=C_m \cup C_u)$ of $D$ with $C\subseteq W$ and $C_u\neq \emptyset$ by Theorem \ref{crownexists}. This is a contradiction to that $D$ is reduced.
\end{proof}

Proceeding from what has been discussed above, we give a polynomial time algorithm which computes a quadratic kernel for the MinLOB-PBGV problem.

\begin{quote}
  \noindent{\bfseries  KERNELIZATION}\\
 \begin{enumerate}
 \item[1.] Build an out-branching $T$ rooted at $r$ by depth-first search.
 \item[2.] Transform $T$ into a minimal out-branching using 1-change.
 \item[3.] {\bf If} the number of leaves of $T$ is at most $n-k$, return 'YES'.
 \item[4.] {\bf Otherwise} Reduce by Rule 1 if possible. If this is not possible, return the instance (it is irreducible).
 \newline Let $T$ be the new out-branching obtained by the construction in the proof of Lemma \ref{reduction}.
 \newline Transform $T$ into a minimal out-branching using 1-change.
 \newline Go to line 3.
\end{enumerate}
\end{quote}

Step 1-3 take $O(n^2m)$ time by Lemma \ref{boundedvc}. At step 4, we can construct the bipartite graph $B$ in time $O(n^3)$, and $V(B)$ and $E(B)$ are bounded by $n+2k+4k^2=O(n^2)$ and $m+4k^2n=O(n^3)$ respectively. Due to Theorem \ref{crownexists}, in $O(n^5)$ time we can reduce the instance by Rule 1 or declare the instance irreducible. Since the size of an instance is strictly decreased at each step of the reduction, we conclude that the algorithm KERNELIZATION runs in $O(n^6)$ time.

\section{Solving MinLOB-PBGV}\label{solsec}

In order to achieve a better running time we provide an alternative
way of showing the fixed-parameter tractability of the MinLOB-PBGV
problem based on the notion of \emph{tree decomposition}.

A {\em tree decomposition} of an (undirected) graph $G$ is a pair
$(X,U)$ where $U$ is a tree whose vertices we will call {\em nodes}
and $X=\{X_{i}:\ i\in V(U)\}$ is a collection of subsets of $V(G)$
(called {\em bags}) such that
\begin{enumerate}
\item $\bigcup_{i \in V(U)} X_{i} = V(G)$,

\item for each edge $\{v,w\} \in E(G)$, there is an $i\in V(U)$
such that $v,w\in X_{i}$, and

\item for each $v\in V(G)$ the set of nodes $\{ i :\ v \in X_{i}
\}$ form a subtree of $U$.
\end{enumerate}
The {\em width} of a tree decomposition $(\{ X_{i}:\ i \in V(U) \},
U)$ equals $\max_{i \in V(U)} \{|X_{i}| - 1\}$. The {\em treewidth}
of a graph $G$ is the minimum width over all tree decompositions of
$G$. We use the notation $\tw(G)$ to denote the treewidth of a graph
$G$.

By a {\em tree decomposition of a digraph} $D$ we will mean a tree
decomposition of the underlying graph $UG(D)$. Also,
$\tw(D)=\tw(UG(D)).$

\begin{theorem} \label{twd}
There is an polynomial time algorithm that, given an instance $(D,k)$ of the
MinLOB-PBGV problem, either finds a solution or establishes a tree
decomposition of $D$ of width at most $2k-2$.
\end{theorem}

\begin{proof} By Lemma \ref{boundedvc}, there
is a polynomial time algorithm which either finds a solution or specifies a vertex
cover $C$ of $D$ of size at most $2k-2$. Let $I=\{v_1, \dots,
v_s\}=V(D) \setminus C$. Consider a star $U$ with nodes
$x_0,x_1,\ldots ,x_s$ and edges $x_0x_1,x_0x_2,\ldots ,x_0x_s$. Let
$X_0=C$ and $X_i=X_0\cup \{v_i\}$ for $i=1,2,\ldots ,s$ and let
$X_j$ be the bag corresponding to $x_j$ for every $j=0,1,\ldots ,s.$
Observe that $(\{X_0,X_1,\ldots ,X_s\},U)$ is a tree decomposition
of $D$ and its width is at most $2k-2.$  \end{proof}

\2

Theorem \ref{twd} shows that an instance $(D,k)$ of the MinLOB-PBGV
problem can be reduced to another instance with treewidth  $O(k)$.
Using standard dynamic programming techniques we can solve this
instance in time $2^{O(k\log k)}n^{O(1)}$. We can further accelerate the solution procedure using kernelization. If
we first find the kernel and then establish the tree decomposition,
the resulting algorithm will run in time
$2^{O(k\log k)}+n^6$. Now we have the following result.

\begin{theorem}
The MinLOB-PBGV problem can be solved by an additive FPT algorithm
of running time $O(2^{O(k\log k)}+n^6).$
\end{theorem}

\section{Related Problems}\label{outsec}

In this section we consider transformations from MinPC and MaxIOT
introduced in Section \ref{introsec} into MinLOB. We start from
MinPC.

For a digraph $D$, let $\pc(D)$ be the minimum number of
vertex-disjoint directed paths in $D$. We have the following:

\begin{proposition}
\label{minleaf=pc} Let $D=(V,A)$ be a digraph and let $\hat{D}$ be
the digraph obtained from $D$ by adding a new vertex $s$ and all
possible arcs from $s$ to $V$. Then $\pc(D)=\ell_{\min}(\hat{D})$.
\end{proposition}
\begin{proof} Since a collection of $p$ disjoint directed paths in $D$
covering $V(D)$ corresponds to an out-branching of $\hat{D}$ with
$p$ leaves, we have $\pc(D)\ge \ell_{\min}(\hat{D})$. Let $B$ be an
out-branching of $\hat{D}$ with $p$ leaves. We say that a vertex $x$
of $B$ is {\em branching} if $d^+_B(x)>1.$ Consider a maximal
directed path $Q$ of $B$ not containing branching vertices. Observe
that $B-V(Q)$ has $p-1$ leaves. Thus, we can decompose the vertices
of $B$ into $p$ disjoint directed paths. Deleting the vertex $s$
from this collection of paths, we see that $\pc(D)\le
\ell_{\min}(\hat{D})$. Thus,
$\pc(D)=\ell_{\min}(\hat{D})$.\end{proof}

\2

Fixed-parameter tractability of MinLOB-PBGV and Proposition
\ref{minleaf=pc} imply that the parameterized problem $\pc(D)\le
n-k$ is FPT, too.

For a digraph $D$ and a vertex $v$ in $D$, let $D_v$ denote the
subgraph of $D$ obtained from the subgraph of $D$ induced by all
vertices reachable from $v$ by deleting all arcs entering $v$. The
following result allows us to reduce MaxIOT to MinLOB.

\begin{proposition}\label{propNL}
Let $D$ be a digraph and let $S$ be the set of vertices belonging to
all strongly connected components of $D$ without incoming arcs. Let
$B_v$ be an out-branching of $D_v$ of minimum number of leaves, and
let $s$ be a vertex of $S$ such that $\ell_{\min}(B_s)\le
\ell_{\min}(B_v)$ for each $v\in S$. Then $B_s$ is a maximum
internal out-tree of $D$.
\end{proposition}
\begin{proof}
Let $T$ be a solution to MaxIOT for $D$ with maximum possible
number of {\em leaves} and let $r$ be the root of $T$. Observe that
$r\in S$ as otherwise we would be able to extend $T$ to an out-tree
$T'$ with more internal vertices such that the root of $T'$ is in $S$.
Observe also that $T$ is an out-branching of $D_r$ as otherwise we
would be able to extend $T$ to an out-branching $T'$ of $D_r$ such
that $T'$ has more either leaves or internal vertices than $T$. Clearly,
$\ell_{\min}(B_r)\le \ell_{\min}(B_v)$ for each $v\in S$.
\end{proof}

Together with the above-proved results, Proposition \ref{propNL} implies
that MaxIOT for acyclic digraphs is polynomially-time solvable and
that the problem of finding an out-tree with at least $k$ internal vertices
in an arbitrary digraph $D$ is FPT. Recall that the problem of finding an out-branching with at least $k$ internal vertices has a quadratic kernel. However, the problem of finding an out-tree with at least $k$ internal vertices does not have a polynomial-size kernel unless PH=$\Sigma_{p}^3$. This easily follows from
Lemmas 1-3 in \cite{bodlaender}.

\section{Further Research}\label{ressec}

We have proved that MinLOB-PBGV is FPT. It would be interesting to
check whether MinLOB-PBGV admits significantly more efficient FPT
algorithms, i.e., algorithms of complexity $O(c^kn^{O(1)})$, where
$c$ is a constant. Another interesting question is whether MinLOB-PBGV admits a linear-size kernel or not.

\2

\2

{\bf Acknowledgements}   Research of Gutin and Kim was supported in
part by an EPSRC grant. Part of the paper was written when Razgon
was visiting Department of Computer Science, Royal Holloway,
University of London. The research of Razgon at the Department of
Computer Science, University College Cork was supported by Science
Foundation Ireland Grant 05/IN/I886.

\end{document}